\RequirePackage[2020-02-02]{latexrelease}
\documentclass[twocolumn,english,aps,10pt,superscriptaddress]{revtex4}
\usepackage[T1]{fontenc}
\usepackage{amsmath,graphicx,amssymb,epsfig,babel,dsfont,mathrsfs,color,amsbsy}

\def\bbm[#1]{\mbox{\boldmath $#1$}}

\begin{document}

\title{Electron tunneling induced thermoelectric effects}

\author{Mauricio G\'omez Viloria}
\affiliation{Laboratoire Charles Fabry, UMR 8501, Institut d'Optique, CNRS, Universit\'{e} Paris-Saclay,
2 Avenue Augustin Fresnel, 91127 Palaiseau Cedex, France}

\author{Riccardo Messina}
\email{riccardo.messina@institutoptique.fr} 
\affiliation{Laboratoire Charles Fabry, UMR 8501, Institut d'Optique, CNRS, Universit\'{e} Paris-Saclay,
2 Avenue Augustin Fresnel, 91127 Palaiseau Cedex, France}

\author{Philippe Ben-Abdallah}
\email{pba@institutoptique.fr} 
\affiliation{Laboratoire Charles Fabry, UMR 8501, Institut d'Optique, CNRS, Universit\'{e} Paris-Saclay,
2 Avenue Augustin Fresnel, 91127 Palaiseau Cedex, France}



\begin{abstract}
We introduce a direct (Seebeck) and inverse (Peltier) thermoelectric effect induced by electron tunneling between closely separated conducting films. When a transverse temperature gradient is applied along one of two films, a bias voltage is induced in the second thanks to the heat transfer mediated by electron tunneling through the separation gap. We highlight a non trivial behavior for this Seebeck effect with respect to geometric characteristics of interacting films. Conversely, when an electric current passes through one of two films a strong thermal power can be removed from or inserted in the second film through an induced Peltier effect. In particular we highlight conditions where the induced Seebeck and Peltier coefficients are larger than in the bulk. These induced thermoelectric effects could find broad applications in the fields of energy conversion and cooling at nanoscale. 
\end{abstract}

\maketitle

\section{Introduction}
Thermoelectric effects are the direct conversion of the temperature difference inside a material into a bias voltage (Seebeck effect) and the conversion of an electric current flowing through this material into a negative or positive thermal power (Peltier effect). The efficiency of a thermoelectric system can be evaluated with the figure of merit $ZT$ of material which is defined as $ZT=S_0^2\sigma T/\kappa_T$, where $T$ is the temperature, $S_0=-\Delta \phi/\Delta T$ is the Seebeck coefficient which quantifies the voltage $\Delta \phi$ generated under a temperature difference $\Delta T$, and $\sigma$ and $\kappa_T$ denoting the electrical and thermal conductivity of material, respectively. Recent works~\cite{Majumdar,Rodgers} have been performed using nanomaterials or nanocomposite structures to increase the $ZT$ by reducing the thermal conductivity while keeping its electric counterpart constant. 

In this Letter we introduce thermoelectric effects induced by the tunneling of electrons between two coupled conductors separated by a small gap. When the size of this gap is sufficiently small and a temperature difference is applied along one of two conductors, a spatially varying density of free charges can be induced across the gap by tunneling effect giving rise to a drift and a diffusion current into the second conductor. On the opposite, when a bias voltage is applied along one of the conductors, a heat flux can be extracted from or inserted into the second conductor. We develop a general theory to describe these thermoelectric effects between two metallic films and analyze its main characteristics with respect to the separation gap and films thicknesses.

The paper is organized as follows.  In section.~\ref{sec:model}, we introduced the model which describes the thermoelectric coupling between two films  through electronic tunneling and near-field radiative heat transfer. In the next section, we present a simple toy model which predicts the main features of the induced thermoelectric effects. The main numerical results of nonlinear system of coupled equations are given in Sec.~\ref{sec:Seebeck} for the Seebeck induced and in Sec.~\ref{sec:Peltier} the Peltier induced configurations, respectively. We highlight the potential  of  these thermoelectric effects for practical applications and we end this article with concluding remarks.

\section{Thermoelectricity and extreme-near-field heat exchange}
\label{sec:model}
\begin{figure}
\centering
\includegraphics[scale=0.45]{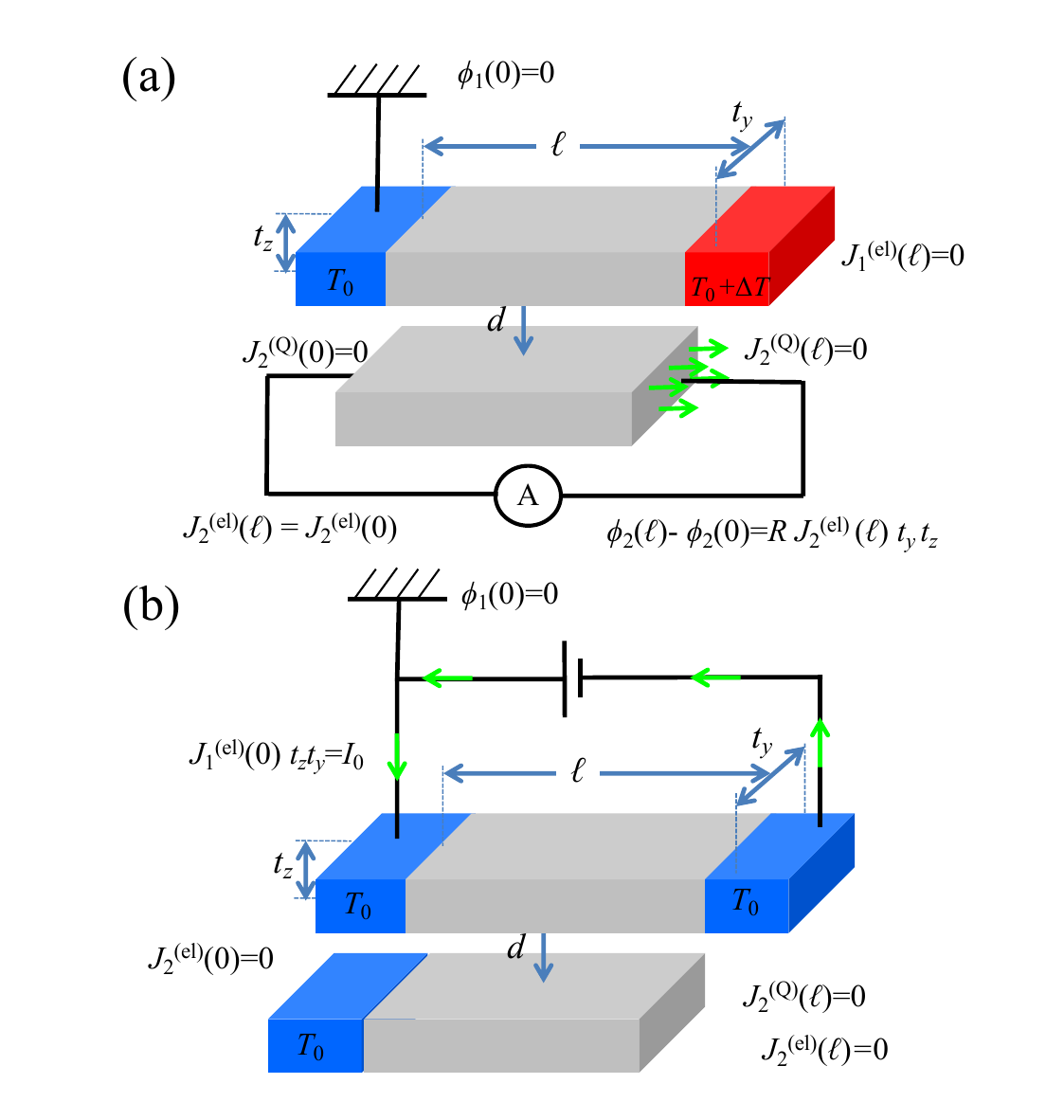}
\caption{\label{fig:1}
Sketch of two conducting films separated by a vacuum gap, coupled via near-field radiative heat exchanges and electron tunneling. (a) Induced Seebeck effect: the top film is connected to two thermal reservoirs at temperatures $T_1(0)$ and $T_1(\ell)$, giving rise to a temperature gradient. It is grounded on its left side [$\phi_1(0)=0$] and no current can escape from its right side [$J^{(\rm el)}_1(\ell)=0$]. The electric current $J^{(\rm el)}_2$ induced in the second film (having adiabatic thermal conditions $J^{(\rm Q)}_2(0)=J^{(\rm Q)}_2(\ell)=0$) can be measured with an ammeter of resistance $R$. (b) Induced Peltier effect: a battery  introduces a current $I_0$ into the top film while it is connected on its two ends to two reservoirs of same temperature $T_1(0)=T_1(\ell)$. The second film is electrically insulated on its two sides [$J^{(\rm el)}_2(0)=J^{(\rm el)}_2(\ell)=0$] and adiabatic conditions are applied on its right side [$J^{(\rm Q)}_2(\ell)=0$] while its left side is connected to a thermal reservoir. The bias voltage induces a thermal power $J^{(\rm Q)}_2(0)$ which can enter or leave the film on its left side.} 
\end{figure}

To start let us consider the system sketched in Fig.~\ref{fig:1} made of two conducting films of same thickness $t_z$, length $\ell$ and width $t_y$, separated by a vacuum gap of thickness $d$.
We assume that along each film ($i=1,2$) small temperature and chemical potential differences $\Delta T_i=T_i(\ell)-T_i(0)$ and $\Delta \mu_i=\mu_i(\ell)-\mu_i(0)$ are applied. Then, according to the Onsager theory~\cite{de Groot}, the local particle current densities $\mathcal{J}_i^{(\rm P)}(x)$ and energy fluxes $\mathcal{J}_i^{(\rm E)}(x)$ along each film are linearly related to the thermodynamic forces $\mathcal{F}_i^{\rm (P)}=\frac{\mathrm d}{\mathrm dx}(\frac{-\mu_i(x)}{T_i(x)})$ and $\mathcal{F}_i^{\rm (E)}=\frac{\mathrm d}{\mathrm dx}( \frac{1}{T_i(x)})$ by the relations
\begin{equation}
\left(\begin{array}{c}
\mathcal{J}_i^{(\rm P)}(x)\\
\mathcal{J}_i^{(\rm E)}(x)
\end{array}\right)=\left(\begin{array}{cc}
L_i^{\rm (PP)} & L_i^{\rm (PE)}\\
L_i^{\rm (EP)} & L_i^{\rm (EE)}
\end{array}\right)\left(\begin{array}{c}
\mathcal{F}_i^{\rm (p)}(x)\\
\mathcal{F}_i^{\rm (E)}(x)
\end{array}\right),\label{Eq:Onsager}
\end{equation}
where $L_i^{(a b)}$ $(a,b=\mathrm P,\mathrm E)$ are the Onsager coefficients which are related to the familiar transport coefficients. Onsager equations can be rewritten in terms of the electric current densities $J^{(\rm el)}_i(x)=-e \mathcal{J}^{\rm (P)}_i(x)$ ($-e$ is the electron charge) and heat fluxes $J^{(\rm Q)}_i(x)=\mathcal{J}^{\rm (E)}_i(x)-\mu(x) \mathcal{J}^{\rm (P)}_i(x)$, and in terms of temperature and bias voltage $\phi_i(x)=-\mu_i(x)/e$ gradients, given by~\cite{pottier} 
\begin{equation}
\label{eq:onsager2}
\left(\begin{array}{c}
J^{(\rm el)}_i(x)\\
J^{(\rm Q)}_i(x)
\end{array}\right)=\frac{\kappa_T}{ {L} }\left(\begin{array}{cc}
1/ T_i(x) & S_0/T_i(x) \\
S_0 &  {L} +S_0^2
\end{array}\right)\left(\begin{array}{c}
-\phi_i'(x)\\
-T_i'(x)
\end{array}\right),
\end{equation}
where we used Ohm's law $J^{(\rm el)}_i(x)=-\sigma \phi_i'(x)$ (at constant temperature), the Seebeck relation, Fourier's law $J^{(\rm Q)}_i(x)=-\kappa_T T_i'(x)$ (at zero current), and Wiedemann-Franz's law $\kappa_T/\sigma= {L}  T_i(x)$ ($ {L} =\pi^2 k^2_{\rm B}/3e^2$ is the Lorenz number, $k_{\rm B}$ is the Boltzmann constant), in order to write the equations in terms of $S_0$ and $\kappa_T$ whose dependence on temperature and bias is negligible for conducting elements near ambient temperatures~\cite{firstprinciples}. Note that when dealing with metals thermal conduction is mainly driven by electrons.

The two films exchange heat by thermal radiation, 
however close to the contact electrons can tunnel through the separation gap carrying both charge and heat. 
A precise understanding of the transition between the radiative and the conduction regime is still an active area of research, both experimentally~\cite{reddy_17,kittel_17} and theoretically~\cite{Francoeur,guo22,Mauricio2}.  The possibility of heat transfer by  tunneling of phonons close to contact has also been suggested~\cite{pendry16}, yet such contributions are predicted to be negligible in metals, except for a specific range of values of the bias voltage~\cite{Mauricio2}.
We restrict our discussion to the electronic and radiative contribution.
 In the steady-state regime, the corresponding energy and charge conservation equations for a given volume element (of each body $i=1,2$) read
\begin{equation}\label{eq:coupling}
\begin{split}
	-t_z \frac{\text{d}J^{(\rm Q)}_i(x)}{\text{d}x}+(-1)^i\Phi^{\rm (rad)}(x)+\Phi^{\rm (el)}_i(x)&=0,\\
	-t_z \frac{\text{d}J^{(\rm el)}_i(x)}{\text{d}x}+(-1)^iJ^{\rm (tun)}(x)&=0,\\
\end{split}
\end{equation}
where $(-1)^i\Phi^{\rm (rad)}(x)$ and $\Phi^{\rm (el)}_i(x)$ represent the radiative and electronic heat fluxes on body $i$ (defined below), whereas $J^{\rm (tun)}(x)$ is the tunneling current density, defined as~\cite{Simmons}
\begin{align}
\label{eq:current}
	J^{\rm (tun)}(x)&=\frac{e m_e}{2\pi^2 \hbar^3}\int_{\tilde{E}(x)}^{\infty}\mathrm{d}E_z\int_0^{\infty}\mathrm{d}E_{\perp}\\
&\times\Delta n_{\rm FD}(E,T_1(x),T_2(x),\mu_1(x),\mu_2(x)) \mathcal{T}^{\rm (el)}(E_z),\nonumber 
\end{align}
where $m_e$ is the electron mass, $E=E_\perp+E_z$ its total kinetic energy decomposed in contributions stemming from velocities perpendicular and parallel to the exchange surface, the bottom of the integral goes from the bottom of the local band $\tilde E(x)=\mathrm{max}\{0,-e[\phi_2(x)-\phi_1(x)]\}$, and $\Delta n_{\rm FD}$ is the difference of Fermi-Dirac distributions, depending on both local temperature $T_i$ and local chemical potential $\mu_i$ associated with each medium. In Eq.~\eqref{eq:current} $\mathcal{T}^{\rm (el)}(E_z)$ is the electronic transmission probability
to overcome the electronic barrier induced between the metals and vacuum gap. In this paper we model this barrier as the solution of a nonlocal Poisson equation~\cite{sidyakin}, calculated analytically using the specular reflection approximation and the Thomas-Fermi approximation for the dielectric function, as done in Refs.~\cite{ilchenko,Mauricio2,Mauricio}. In order to recover the transmission probability through the barrier, we adopt an $S$-matrix algorithm as employed in Ref.~\cite{Mauricio}.

Electron tunneling also gives rise to a heat transfer which takes the form~\cite{Xu,Mauricio2}
\begin{align}
\label{eq:flux_el}
	&\Phi_i^{(\mathrm{el})}(T_1,T_2,d,x)=(-1)^i\frac{ m_e}{2\pi^2 \hbar^3}\int_{\tilde E}^{\infty}\mathrm{d}E_z\int_0^{\infty}\mathrm{d}E_{\perp}\\
	\nonumber&\times (E-\mu_i(x))\Delta n_{\rm FD}(E,T_1(x),T_2(x),\mu_1(x),\mu_2(x) \mathcal{T}^{\rm (el)}(E_z).
\end{align}

Analogously, the flux carried by photons, evaluated from fluctuational-electrodynamics theory~\cite{Polder,Joulain_rev,Volokitin_rev, RMP}, reads
\begin{align}
\label{eq:Flux_D}
\Phi^\text{(rad)}(x)=&\underset{l=\mathrm s,\mathrm p}{\sum}\int_0^\infty\!\frac{\mathrm d\omega}{2\pi}\ \hbar\omega \Delta n_{\rm BE}(T_1(x),T_2(x))\\
\nonumber&\times\int_{0}^{\infty} \frac{\mathrm d\kappa}{2 \pi} \kappa\, \mathcal{T}_l^\text{(rad)}(\omega,\kappa), 
\end{align}
where $\mathcal{T}_l^\text{(rad)}(\omega,\kappa)$ denotes the energy transmission coefficient for one electromagnetic mode $(\omega,\kappa)$ for the polarizations $l = \mathrm s,\mathrm p$,
 $n_{\rm BE}(\omega,T)=1/[\exp(\hbar\omega/k_{\rm B}T)-1]$ is the Bose-Einstein distribution function, and
\begin{equation}\begin{split}
	\label{eq:transmission_rad}
	&\mathcal{T}^{\rm (rad)}_{l}(\omega,\kappa)\\
	&\,=\begin{cases} 
		\displaystyle\frac{(1-|r_{l}|^2)^2}{|1-r_{l}^2\exp(2\mathrm i k_z d )|^2}, & k<\omega/c,\\\vspace{-0.3cm}\\
		\displaystyle\frac{4\,(\mathrm{Im}\,r_{l})^2\exp(-2\,\mathrm{Im}\,k_z d)}{|1-r^2_{l}\exp(-2\,\mathrm{Im}\,k_z d )|^2}, & k\geq \omega/c,
	\end{cases}
\end{split}\end{equation}
is the radiative transmission probability, where $k_z=\sqrt{(\omega/c)^2-\kappa^2}$. and the reflection coefficients are given by Fresnel's formulas,
\begin{equation}
\label{eq:r_s}
	r_{\mathrm s}(k,\omega)=\frac{k_z-k_{\mathrm m, z}}{k_z+k_{\mathrm m, z}},\quad r_{\mathrm p}(k,\omega)=\frac{\epsilon(\omega) k_z-k_{\mathrm m, z}}{\epsilon(\omega) k_z+k_{\mathrm m, z}},
\end{equation}
where $k_{\mathrm m,z}=\sqrt{ (\omega/c)^2 \epsilon(\omega)-\kappa^2}$ is the $z$ component of the wavevector inside the media. To describe metals, we employ a local description of the dielectric susceptibility from the Drude model, given by	$\epsilon(\omega)=1- \omega_{\rm pl}^2/\omega(\omega+\mathrm i \Gamma)$, where $\omega_{\rm pl}$ is the plasma frequency of the metal, and $\Gamma$ is the damping coefficient. 

To analyze the thermoelectric effects induced by electron tunneling we solve the nonlinear system of Eqs. (\ref{eq:onsager2}) and (\ref{eq:coupling}) with respect to the temperatures $T_i(x)$ and bias voltage $\phi_i(x)$ profiles in the case of two complementary scenarios, corresponding to suitable configurations where the direct (Seebeck) and inverse (Peltier) thermoelectric effects can be observed. To this end, we first replace the expression of tunnel current by its linearized form in terms of both $T_i(x)$ and $\phi_i(x)$, which yields
\begin{equation}
J^{\rm (tun)}(x)\approx G_J(d)[\phi_1(x)-\phi_2(x)]-h_J(d)[T_1(x)-T_2(x)],\label{eq:heat_linear}
\end{equation}
where $G_J$ and $h_J$ denote, respectively, the electrical and thermal tunneling conductances (per unit surface) that depend only on the gap thickness $d$. On the other hand, the heat fluxes must be expanded to the quadratic order
\begin{equation}
\begin{split}
	\Phi^{\rm (el)}_1(x)&\simeq G_\Phi(d) [\phi_1(x)-\phi_2(x)] + \frac{G_J(d)}{2} [\phi_1(x)-\phi_2(x)]^2\\ 
	 &- h_\Phi(d)[T_1(x) - T_2(x)]	 -\tilde{h}_\Phi (d)[T_1(x) - T_2(x)]^2\\
	  &- \frac{h_J(d)}{2}[\phi_1(x)-\phi_2(x)][T_1(x) - T_2(x)]\\
	\Phi^{\rm (el)}_2(x)&\simeq -G_\Phi(d) [\phi_1(x)-\phi_2(x)]+ \frac{G_J(d)}{2} [\phi_1(x)-\phi_2(x)]^2\\
	& +h_\Phi(d)(T_1 - T_2)	+\tilde{h}_\Phi(d)[T_1(x) - T_2(x)]^2\\ 
	&- \frac{h_J(d)}{2}[\phi_1(x)-\phi_2(x)][T_1(x) - T_2(x)],\label{eq:elec_quad}
\end{split}
\end{equation}
so that the tunneling current density can satisfy the energy conservation equation
\begin{equation}
\Phi_1^{\rm (el)}(x)+\Phi_2^{\rm (el)}(x)=J^{\rm (tun)}(x)[\phi_1(x)-\phi_2(x)].\label{eq:elec_power}
\end{equation}
The left-hand side in this expression corresponds to the thermal power mediated by electron tunneling through the separation gap while the right-hand side is the electric power associated to tunnel current. In Eq.~\eqref{eq:elec_quad}, $G_\Phi(d)$, $h_\Phi(d)$ and $\tilde{h}_{\Phi}(d)$, are the thermal conductances and Hessian associated to the bias voltage and temperature differences. 
For short distances, all parameters behave as an exponential of the form $G_J(d)=G_{J0}\exp[- k_1 d]$, $G_\Phi(d)=G_{\Phi0}\exp[- k_2 d]$, $h_J(d)=h_{J0}\exp[- k_3 d]$ and $h_\Phi(d)=h_{\Phi0}\exp[- k_4 d]$, where $G_{J0},G_{\Phi0},h_{J0},h_{\Phi 0},k_1,k_2,k_3,k_4$ are constants, obtained from a fit from the general expressions of Eqs.~\eqref{eq:current}, \eqref{eq:flux_el} and \eqref{eq:Flux_D}. 

\section{Linear four-node thermoelectric circuits}
\label{sec:toymodel}
\begin{figure*}[t]
    \centering
    \includegraphics[width=0.9\textwidth]{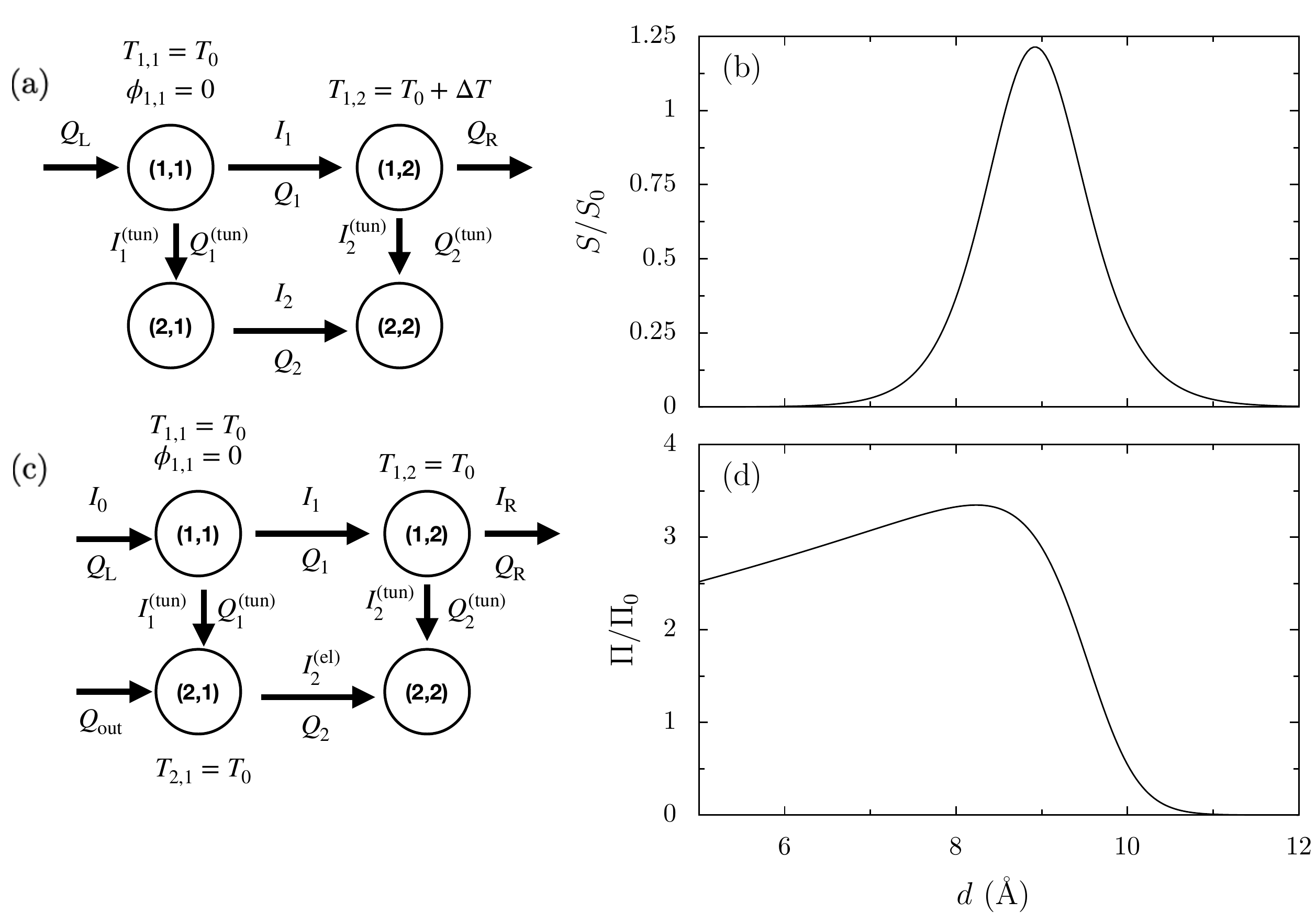}
    \caption{ Induced thermoelectric coefficients as a function of the separationg gap $d$ for the simplified model, simplified Seebeck circuit in panel (a) and simplified Peltier circuit in panel (c). In panel (b) the induced Seebeck coefficient $S$ is shown (in units of $S_0$) for $t_z=5\,\mu$m.  In panel (b) the induced Peltier coefficient $\Pi$ (in units of $\Pi_0=S_0T_0$) is shown for $t_z=1\,\mu$m. For both cases, we set $\ell=t_y=1\,$mm and $T_0=300\,$K.}
    \label{fig:diagram}
\end{figure*}

Before discussing the solutions to the coupled equations, we provide a simplified linear circuit model in order to introduce the main features of the induced thermoelectric coefficients. We consider the limit of small $\Delta T$ and $\Delta\phi$ in order to obtain a linear system of equations, and we simplify the system to two 4-node thermoelectric circuits as shown in Fig.~\ref{fig:diagram}(a) (simplified Seebeck) and in Fig.~\ref{fig:diagram}(c) (simplified Peltier), which are linear analogs of Fig.~\ref{fig:1}(a) and Fig.~\ref{fig:1}(b), respectively.
For the horizontal fluxes, we consider only the diagonal terms in the Onsager equations in Eq.~\eqref{eq:onsager2}, which yields
\begin{equation}
\begin{cases}
	\displaystyle Q_{i} =-\kappa_T\left(1+\frac{S_0^2}{L}\right) (T_{i,2}-T_{i,1})t_yt_z\\
	I_i  = -\sigma(\phi_{i,2}-\phi_{i,1})t_yt_z
	\end{cases}
\end{equation}
where $Q_i$ ($I_i$) is the heat (electric current) going from node $(i,1)$ to node $(i,2)$, where $i,j=1,2$,  in analogy to the conduction in the metallic films. $\phi_{i,j}$ and $T_{i,j}$ are the temperature and electric potentials of nodes $(i,j)$, respectively. Similarly, in analogy to Eq.~\eqref{eq:coupling} for the vertical edges, we have 
\begin{equation}\label{eq:fluxes}
\left\{
\begin{matrix}
	 Q_j^{\rm (tun)} =\frac12 [G_\Phi(d)\,(\phi_{2,j}-\phi_{1,j}) - h_\Phi(d)\,\,(T_{2,j}-T_{1,j})]t_y\ell \\
	 I^{\rm (tun)}_j=\frac12 [G_J(d)\,(\phi_{2,j}-\phi_{1,j}) - h_J(d)\,(T_{2,j}-T_{1,j}) ]t_y\ell
\end{matrix}
\right.
\end{equation}
where $Q_j^{\rm tun}$ ($I_j^{\rm tun}$) is the heat (electric current) carried by the tunneling electrons going from node $(1,j)$ into $(2,j)$. The thermoelectric couplings are given by the tunneling fluxes in this simplified system. For this simplified discussion, we neglect $G_\Phi(d)$ for the circuit of Fig.~\ref{fig:diagram}(a), and reciprocally $h_J(d)$ for the the configuration in Fig.~\ref{fig:diagram}(c).

By imposing the conservation of heat and electric currents,  we can calculate the induced Seebeck coefficient $S=-(\phi_{2,2}-\phi_{2,1})/(T_{1,2}-T_{1,1})$ for the simplified Seebeck circuit, and the induced Peltier coefficient $\Pi=Q_{\rm out}/I_0$ for the simplified Peltier circuit, where $Q_{\rm out}$ is the heat entering the second layer and $I_0$ the current imposed in the first layer as shown in Fig.~\ref{fig:diagram}(c). By solving the continuity equations analytically, we obtain
\begin{equation}\label{eq:seebeck_simple}
	S=\frac{2 h_J(d) \kappa_T (1+S^2_0/L) t_z \ell}{[2\sigma t_z  +G_J(d)  \ell][4\kappa_T(1+S^2_0/L)t_z +h_{\Phi}(d) \ell]}\,,
\end{equation}
for the simplified Seebeck configuration, plotted in Fig.~\ref{fig:diagram}(b), and for the simplified Peltier configuration, we find 
\begin{equation}\label{eq:peltier_simple}
	\Pi=\frac{  h_\Phi(d) G_{\Phi}(d) \ell^2 }{[2\sigma t_z  +G_J(d) \ell][4\kappa_T (1+S_0^2/L)t_z+2h_\Phi(d)  \ell]}\,,
\end{equation}
plotted in Fig.~\ref{fig:diagram}(d).

We observe that this simplified circuit has two interesting behaviours, first it predicts thermoelectric coefficients that are larger than the bulk at a given distance, and secondly, the coefficients behave non-monotonically as a function of distance. The origin of the increased values comes from the vertical coupling to the vacuum that plays the role of a negative Seebeck element, of opposite sign to that of noble metals. Such thermoelectric effects that surpass classical values could hold significant promise for thermal management in reduced systems. We have verified that this behaviour is very robust to the introduction of the other terms in Eq.~\eqref{eq:onsager2}, however the peak values and the position of the peaks of the thermoelectric coefficients will depend on the shift of the nonlinearities of the full system, as shown in the following section.

The non-monoticity of Fig.~\ref{fig:diagram}(b) and  Fig.~\ref{fig:diagram}(d) is the result of the denominators in Eq.~\eqref{eq:seebeck_simple} and Eq.~\eqref{eq:peltier_simple}, which are the product of two terms, each one consisting in the sum of two competing terms, the former associated with transport within the film (proportional to $\sigma$ or $\kappa$), the latter to tunneling between the films (containing one of the distance-dependent conductances $G_J(d)$, $h_\Phi(d)$ and $h_J(d)$). While the tunneling terms dominate at short distance, they become negligible at larger distances because of their exponential behavior. This competition, combined with the exponential behavior at the numerator, is at the origin of non-monotonous behaviour.

\section{Results}
\subsection{Induced Seebeck effect}
\label{sec:Seebeck}
In this section, we discuss the results for the full systems of differential equations given by Eq.~\eqref{eq:onsager2} and Eq.~\eqref{eq:coupling}, for two scenarios in Fig.~\ref{fig:1}. We also provide an explicit description of the different boundary conditions.
In the first scenario, [see Fig.~\ref{fig:1}(a)], a primary temperature gradient is applied along body 1 [i.e. $T_1(0)=T_0$ and $T_1(\ell)=T_0+\Delta T$] and this body is electrically connected to the ground at $x=0$ [i.e. $\phi_1(0)=0$] and no electric current can escape from its opposite side [i.e. $J^{(\rm el)}_1(\ell)=0$]. As for the second, we assume it is connected to an ammeter of resistance $R$ and a current is free to circulate through it thanks to the primary temperature gradient in body 1. Therefore, according to Ohm's law, the bias voltage $\Delta\phi_2=\phi_2(\ell)-\phi_2(0)$ along body 2 satisfies $\Delta\phi_2=R\:J^{(\rm el)}_2(\ell)\:t_zt_y$. We assume adiabatic conditions at both ends of body 2 [$J^{(\rm Q)}_2(0)=J^{(\rm Q)}_2(\ell)=0$] and continuity of the current through the resistance [i.e. $J^{(\rm el)}_2(0)=J^{(\rm el)}_2(\ell)$]. The differential system \eqref{eq:coupling} associated with expressions \eqref{eq:elec_quad} and \eqref{eq:elec_power} is solved numerically using a fourth order collocation method as described in Ref.~\cite{diffequations}. The induced electromotive force that develops in the second solid when a temperature gradient is applied in the first one can be quantified by the induced Seebeck coefficient, $S=-\Delta \phi_2/\Delta T_1$, directly proportional to the current $I_2(\ell)$ [$I_i(\ell)=J^{(\rm el)}_i(\ell)t_zt_y$] via Ohm's law.

\begin{figure}
\centering
\includegraphics[width=0.5\textwidth]{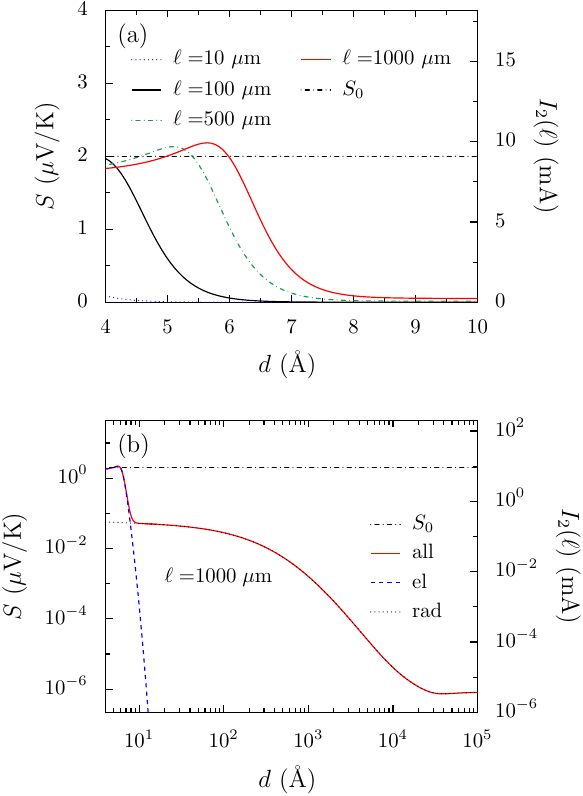}
\caption{\label{fig:2}
Induced Seebeck coefficient $S$ and induced current $I_2(\ell)$ as a function of the gap thickness $d$ between two gold films of thickness $t_z=5\:\mu$m. The first film is held at $T_0=300\;$K on its left end and a temperature difference $\Delta T=100\;$K is applied along it. The second film is connected to a resistance of $R=22\;$m$\Omega$. In (a) and (b) the Seebeck coefficient for bulk gold $S_0\approx 2\;\mu$m \cite{seebeckgold} is represented by a dot-dashed line. In (a) the curves are plotted for various values of the system length $\ell$, in (b) for a large range of distances (red). In (b) the induced Seebeck coefficient due to the electronic contribution (blue dashed line) and the purely radiative contribution (black dotted) are also shown. 
} 
\end{figure}

The results are plotted in Fig.~\ref{fig:2} for two gold films ($t_y=\ell$) as a function of their separation distance.
In Fig.~\ref{fig:2}(a) we see that the induced Seebeck coefficient increases at close separation distances (as a result of an increase tunneling of free charges), and also when the length of the system increases. We stress that our mesoscopic approach is expected to fail when approaching contact ($d$ of the order of the lattice constant), when $S$ is expected to tend to its bulk value $S_0$. It is worthwhile noting that for sufficiently long films, this coefficient can go beyond the bulk Seebeck coefficient $S_0$ [dot-dashed line in Fig.~\ref{fig:2}(a)]. 
As discussed in Sec.~\ref{sec:toymodel}, the non-monotonous dependence of the $S$ as function of distance is a result of the competition between thermal and electric conduction coefficients.

 In Fig.~\ref{fig:2}(b) we show the evolution of this effect at larger separation distances and highlight the relative contributions of different carriers. For distances larger than 1\:nm the electronic contribution to the transfer falls down very rapidly and the coupling between the two films is manly due to, first, non-propagative thermal photons (i.e. near-field radiative heat transfer) for separation distances smaller than the thermal wavelength and to propagative photons (i.e. far field transfer) at larger distances. For large distances, the temperature profile in body 1 becomes linear $T_1(x)=T_0+x\Delta T /\ell$, the two films being almost uncoupled and the temperature in body 2 reaches a constant temperature of $T_2\simeq T_0+\Delta T/2$, as shown in Appendix~\ref{app:profiles}. It is interesting to note that the dependence of the induced Seebeck coefficient with respect to the gap thickness could be used to develop a metrology of near-field and even extreme near-field heat exchanges.

\begin{figure}
\centering
\includegraphics[width=0.5\textwidth]{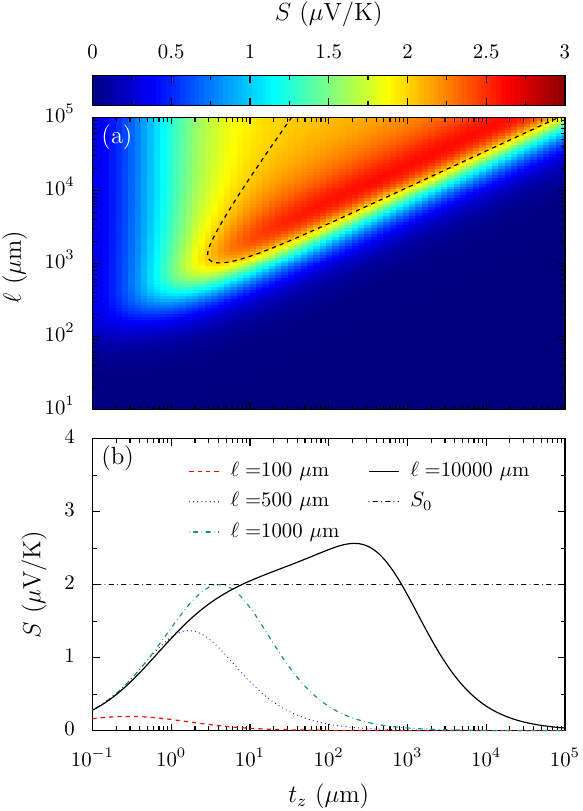}
\caption{\label{fig:3}
Induced thermopower between two gold films ($t_y=\ell$) separated by a gap distance $d=6\;$\AA{} as a function of the film thickness $t_z$, held at $T_0=300\;$K and $\Delta T=100\;$K, and a resistance of $R=22\;$m$\Omega$. In (a) the Seebeck coefficient is represented in a density plot as function of $\ell$ and $t_z$. Dot-dashed line indicates the value of $S_0$. In (b) we represent two dimensional cuts of (a) for different $\ell$. The dot-dashed line indicates the Seebeck coefficient $S_0$ of gold (bulk).} 
\end{figure}

We also analyze the dependence of the induced Seebeck coefficient on the film thickness and length. In Fig.~\ref{fig:3}(a) we observe that there is an intermediate region in the space of parameters $\ell$ and $t_z$ where $S$ goes beyond $S_0$ and it can even exceed it by almost 50\%. We have verified numerically that for large $\ell$ and $t_z$ the contour lines of constant $S$, including its maximum value, have the form $t_z\propto\ell^2$.
 In Fig.~\ref{fig:3}(b) we note, at a given length, that $S$ is a non-monotonic function of $t_z$: more specifically, the induced Seebeck coefficient vanishes for both infinitely thin and thick films. For ultra-thin thicknesses the coupling between the two films tends to vanish and only a residual current can flow inside body 2. On the opposite, for thicker films, most of the thermal power is carried by conduction inside each film so that the coupling induced by electron tunneling or radiative transfer has little impact of the induced current.

\subsection{Induced Peltier effect}
\label{sec:Peltier}
Let us now investigate the inverse induced thermoelectric effect. In this case we consider the configuration as depicted in Fig.~\ref{fig:1}(b). A bias voltage [grounded to the left $\phi(0)=0$] is applied along body 1 in order to introduce a current $I_0=J_1(0)t_zt_y$ and its two ends are coupled to a thermostat at the same temperature [$T_1(0)=T_1(\ell)=T_0$]. As the second body is concerned, it is also coupled to the same thermostat on its left side [$T_2(0)=T_0$] and thermally insulated on its opposite side [$J^{(\rm Q)}_2(\ell)=0$] for simplicity. Moreover, this film is an open circuit so that no electric current can flow through it [$J^{(\rm el)}_2(0)=J^{(\rm el)}_2(\ell)=0$]. We show that a current in body 1 can induce a thermal current in body 2. The heating or cooling which can be induced in the second film when an electric current flow in the first one can be quantified by the induced Peltier coefficient $\Pi=J^{(\rm Q)}_2(0)/J^{(\rm el)}_1(0)$. We remind that for a single body at temperature $T_0$, this coefficient satisfies $\Pi_0=S_0\:T_0$. 

\begin{figure}
\centering
\includegraphics[width=0.5\textwidth]{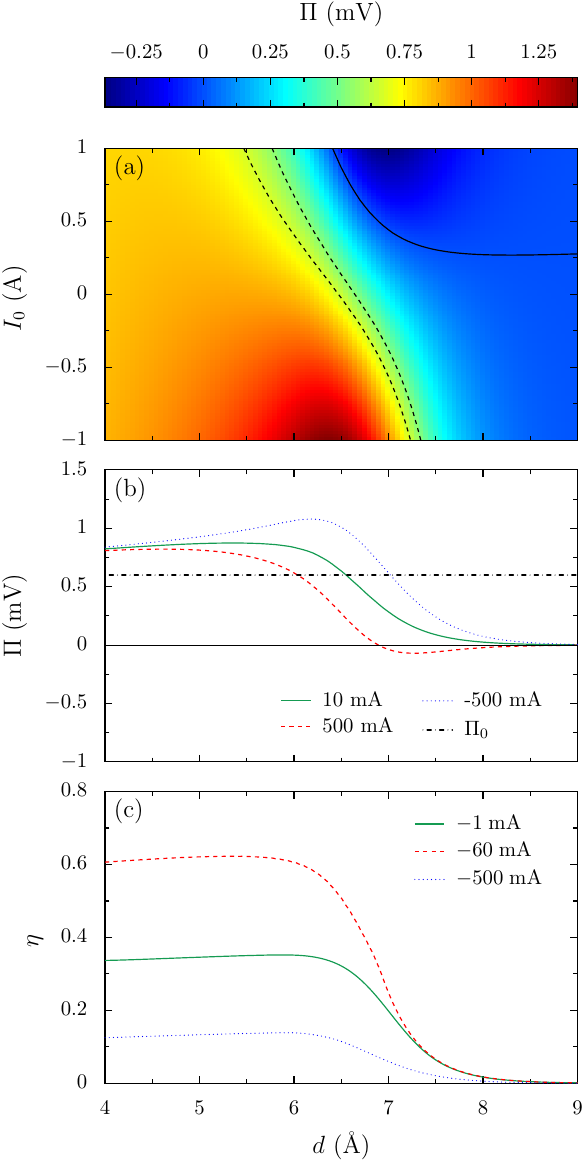}
\caption{\label{fig:4}
Induced Peltier coefficient and cooling efficiency as a function of $d$ for two gold films ($t_y=\ell$) of $t_z=1\;\mu$m, $\ell=1000\;\mu$m and $T_0=300\;$K. (a) Induced Peltier coefficients as a function of the current $I_0$ applied in body 1 and $d$. Dashed lines correspond to $\Pi_0=S_0T_0$ of gold $\pm10\%$, and the solid line represents $\Pi=0$. Cuts of (a) are shown in panel (b) for $I_0=10\,$mA (solid line), 500 mA (dashed line) and -500\;mA (dotted line). The black dotted line represents $\Pi_0$. (c) Cooling efficiency as a function of $d$ for $I_0=-1\;$mA (solid line), $-500\;$mA (dotted line) and $-60\;$mA (dashed line).
} 
\end{figure}

In Fig.~\ref{fig:4}(a) we show the evolution of $\Pi$ with respect to the gap thickness and intensity of electric current flowing through body 1.
When the current is negative, i.e. goes from right to left, $\Pi$ remains positive and asymptotically vanishes for large $d$. As for the induced Seebeck effect, the induced Peltier effect can exceed the bulk value $\Pi_0$ at short range distances thanks to electron tunneling. This induced effect can also change sign. Hence $\Pi$ can become negative for distances smaller than 1\;nm and the electric current $I_0$ is positive (from left to right). In this case the Peltier effect heats up the second solid. Moreover we see in Fig.~\ref{fig:4}(b) that above 8\;\AA{} the induced Peltier coefficient rapidly decreases whatever the electric current flowing through body 1. However, below this threshold, the it has a strong dependence on the current and can go beyond $2\Pi_0$ for $d=6.5\:$\AA{} when a current $I_0=-500\;$mA is flowing in body 1. This corresponds to an extracted power through the left side of body 2 of $50\:$W/cm$^2\times t_z\ell$. This flux is larger than most of thermoelectric cooling devices.

To conclude we analyze the cooling efficiency of induced Peltier effect, defined as
\begin{equation}
\label{eq:eta}
	\eta=-\frac{J^{\rm (Q,-)}_2(0)}{J_1^{\rm (Q,+)}(0)-J^{\rm (Q,-)}_1(\ell)-J^{(\rm el)}_1(0)\Delta\phi_1},
\end{equation}
where $-J^{(\rm el)}_1(0)\Delta\phi_1\geq 0$ is the electrical power density injected by the battery, and $J^{\rm (Q,+)}_i(x)=\operatorname{max}\{J^{(\rm Q)}_i(x),0\}$ and $J^{\rm (Q,-)}_i(x)=\operatorname{min}\{J^{(\rm Q)}_i(x),0\}$ are to be accounted when there is heat entering the system through the thermal reservoirs. In Fig.~\ref{fig:4}(c) we represent this efficiency for different values of current applied in film 1. We see that $\eta$ reaches a maximum at a given value of the current [about $I_0=-60\;$mA], and it decreases for smaller and larger values. The cooling efficiency can reach values of about 60\%, an efficiency comparable and even greater than that of current thermoelectric cooling systems~\cite{Riffat} and of conventional Rankine systems based on compression/expansion cycles~\cite{Kaushikt}.

\section{Concluding remarks}
\label{sec:conclusion}

In summary, we have highlighted two thermoelectric effects induced by electron tunneling between two closely separated conducting films. We have demonstrated that the induced Seebeck and Peltier effects can exceed the classical thermoelectric properties of bulk material when the films are separated by subnanometric gaps. We have shown that the induced Seebeck effect is very sensitive to the separation distance, making the measurement of the induced thermopower a promising method to quantify the extreme-near-field transfer between two conductors. Beside this effect we have highlighted the conditions to extract thermal power from a solid using the Peltier effect induced by electron tunneling and shown that the performances of this effect are comparable and even better than the classical compression-cycle systems. These thermoelectric effects could be exploited for nanoscale energy conversion and to develop solid-state refrigeration devices.

\begin{acknowledgments}
P.B.-A. acknowledges fruitful and inspiring discussions with Janet Treger. We acknowledge financial support from Labex Nanosaclay, ANR-10-LABX-0035 (Flagship Project MaCaCQu).
\end{acknowledgments}

\appendix
\section{Induced-Seebeck profiles}
\label{app:profiles}
In this Appendix, we illustrate the behaviour of temperature, electric potentials and current profiles along the film in the $x$ direction, as a function of the separation gap $d$ for the induced Seebeck configuration, as shown in Fig.~\ref{fig:1}(a). For this purpose, we choose three different regimes:   $d=6.5\,$\AA{} ($S<S_0$),  $d=5.5\,$\AA{} ($S>S_0$) and  $d=4.5\,$\AA{} ($S<S_0$); based on the non-monotonous dependence of the induced Seebeck coefficient $S$ as a function of $d$ as shown in Fig.~\ref{fig:2}(a) (for $\ell=1000\,\mu$m). In the following discussion we also fix $t_y=\ell$ and $t_z=5\,\mu$m for the dimensions of the films, $T_1(0)=300\,$K and $T_1(\ell)=400\,$K for the temperatures of the reservoirs, and $R=22\,$m$\Omega$ for the ammeter resistance.  

\begin{figure}[t]
    \centering 
    \includegraphics[width=0.5\textwidth]{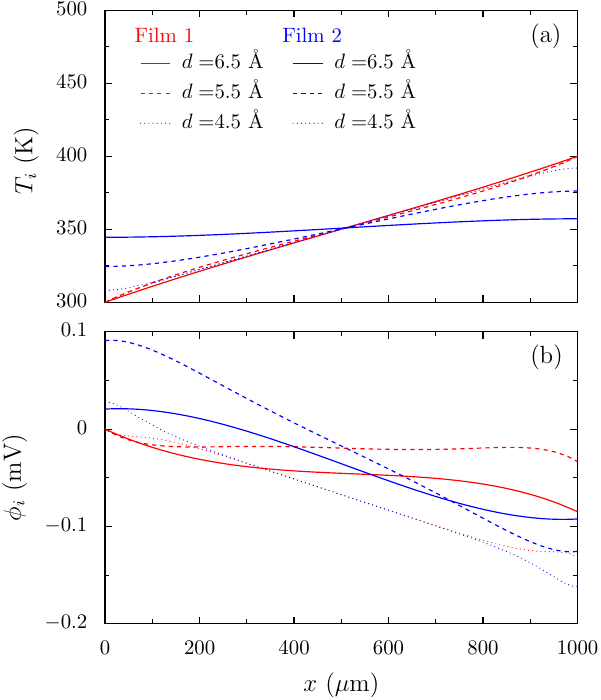}
    \caption{ Profiles for different distances (solid, dashed, dotted). In panel (a) temperature profiles for film 1 (red) and film 2 (in blue). In panel (b) electric potential profiles for films 1 and 2. }
    \label{fig:Tphi}
\end{figure}

For large distances the behavior of film 1 (film 2) tends to the one of an uncoupled film, i.e. zero current and linear (constant) dependence of temperature and electric potential as a function of $x$. In Fig.~\ref{fig:Tphi}(a), we plot the temperature profiles for the three different separation distances.
We observe that $T_1(x)$ remains mostly linear at any separation distance $d$, however the temperature profile $T_2(x)$ is constant (at about $350\,$K) for large distances and approaches $T_1(x)$ as $d$ approaches contact. In Fig.~\ref{fig:Tphi}(b), we plot the electric potential profiles of materials 1 and 2. For large distances $\phi_1(x)$ is linear and it tends to Seebeck's relation $\phi_1(x)+ST(x)\to0$, however its electric potential remains always negative (with respect to the ground) and varies strongly with the separation distance, behaving almost as a constant near the maximum of $S$, before becoming linear again for smaller distances. The electric potential $\phi_2(x)$ can take positive values and near $d=6\,$\AA{} it becomes linear as a function of $x$ and tends to $\phi_1(x)$ near contact. We can observe in Fig.~\ref{fig:Tphi}(b) that for $d=5.5\,$\AA{} the slope is large leading to a significant induced $S$.

\begin{figure}[b]
    \centering
    \includegraphics[width=0.5\textwidth]{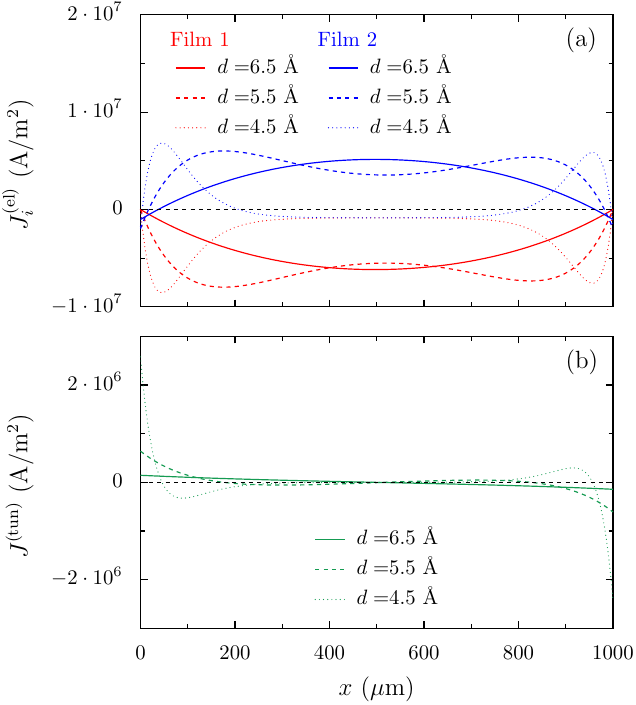}
    \caption{ Current density profiles for different distances (solid, dashed, dotted). In panel (a) in plane current density profiles for film 1 (red) and film 2 (in blue). In panel (b) vertical tunneling density current profile (in green), positive values indicate a current going from material 1 to material 2.}
    \label{fig:JJ}
\end{figure}

The origin of the non-monotonous behavior of the system can also be illustrated by inspecting the electric current densities. In Fig.~\ref{fig:JJ}(a), we plot the in-plane current densities as function of $x$ for the three values of $d$.
For all values of $d$, $J_1^{\rm (el)}(x)$ is negative (going from the right to the left), it is slightly asymmetrical with respect to $x=\ell/2$, and there is no lateral current escaping material 1 at $x=0$ and $x=\ell$. Near contact, the current gets reduced, but shows large variations near the boundaries. Analogously, $J_2^{\rm (el)}(x)$ follows a similar pattern but with the main difference that the current direction can get inverted at several points along the film. These direction flips are also related to the sign flips of the tunneling current density shown in Fig.~\ref{fig:JJ}(b). The tunneling currents are mostly concentrated near the boundaries. Contrary to $J^{\rm (el)}_i(x)$, the tunneling current  $J^{\rm (tun)}(x)$ is antisymmetric with respect to $x=\ell/2$ due to the conservation of the electric charge.

\begin{figure*}[t]
    \centering
    \includegraphics[width=\textwidth]{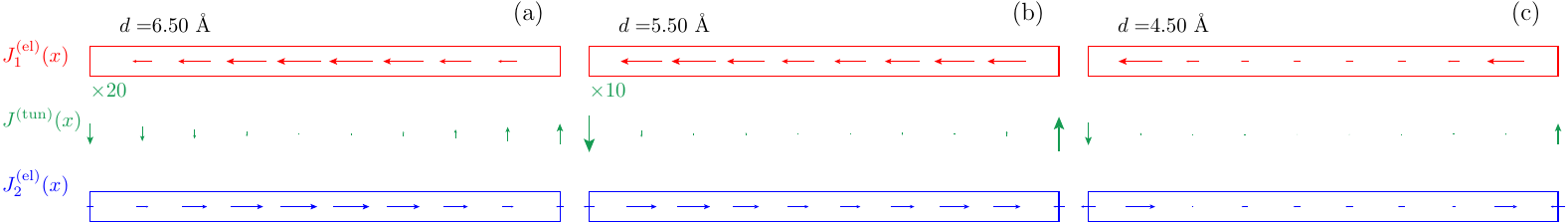}
    \caption{Current density fields across the whole system for three different distances (a) $d=4.5\;$\AA{}, (b) $d=5.5\;$\AA{} (c) $d=6.5\;$\AA{}. The arrows are normalized with respect to the maximum of current density for each system. Tunneling currents in panels (b) and (c) are multiplied by a factor of 10. }
    \label{fig:fields}
\end{figure*}

We end this Appendix by illustrating the current density fields in Fig.~\ref{fig:fields}.
For the largest distances ($d=6.5\,$ and $d=5.5\,$ \AA{}) the currents circulate anti-clockwise, going to the left in material 1, tunneling near the left boundary $x=0$, into material 2, and bifurcating into the in-plane current to the right in material 2 and the current that leaves the system into the ammeter to the left. These two contributions join again at the right boundary ($x=\ell$) and tunnel back into material 1. For $d=4.5\,$\AA{} the situation is complex as there are extra nodes producing anti-clockwise and clockwise subcircuits between materials 1 and 2, but with reduced intensity near the middle of the system. The only dominating currents left produce a localized anticlockwise circulation near the boundaries. This behaviour is tied to the reduction of the $S$ near contact. The profiles in the Peltier configuration are reciprocally analogous to the Seebeck configuration but the non-monotonous behaviour is in the temperatures and heat current densities.

\end{document}